\documentclass[aps,preprint,a4paper,12pt]{revtex4-1}
\usepackage{amsmath}
\usepackage{hyperref}
\usepackage{amsfonts}
\usepackage{amssymb}
\usepackage{graphicx}
\usepackage{latexsym}
\usepackage{color}
\usepackage{booktabs}
\usepackage{dcolumn}
\usepackage{subfigure}
\usepackage{float}
\usepackage{multirow}
\usepackage{ulem}
\usepackage{xcolor}

\def\be#1\ee{\begin{align}#1\end{align}} 
\renewcommand{\Re}{\operatorname{Re}}
\renewcommand{\Im}{\operatorname{Im}}
\def\be{\begin{eqnarray}}
\def\ee{\end{eqnarray}}

\begin{document}
\title{Late-time tails, entropy aspects, and stability of black holes with anisotropic fluids}
\author{B. Cuadros-Melgar}
\email{bertha@usp.br}
\affiliation{Escola de Engenharia de Lorena, Universidade de S\~ao
   Paulo, Estrada Municipal do Campinho S/N, CEP 12602-810, Lorena, SP, Brazil}
\author{R. D. B. Fontana}
\email{rodrigo.fontana@uffs.edu.br}
\affiliation{Universidade Federal da Fronteira Sul, Campus Chapec\'o, CEP 89802-112, SC, Brazil}
\author{Jeferson de Oliveira}
\email{jeferson@gravitacao.org}
\affiliation{Instituto de F\'i­sica, Universidade Federal de Mato Grosso, CEP 78060-900, Cuiab\'a, MT, Brazil}

\begin{abstract}
In this work we consider black holes surrounded by anisotropic
fluids in four dimensions. We first study the causal structure of these
solutions showing some similarities and differences with Reissner-Nordstr\"om-de
Sitter black holes. In addition, we consider scalar perturbations on this
background geometry and compute the corresponding quasinormal
modes. Moreover, we discuss the late-time behavior of the
perturbations finding an interesting new feature, {\it i.e.}, the
presence of a subdominant power-law tail term. Likewise, we compute
the Bekenstein entropy bound and the first semiclassical correction to
the black hole entropy using the brick wall method, showing their
universality. Finally, we also discuss the thermodynamical stability
of the model. 

\end{abstract}

\maketitle

\section{Introduction}\label{intro}

Recently, the LIGO collaboration
\cite{PhysRevLett.116.061102}\cite{TheLIGOScientific:2016src} started
the age of gravitational wave astronomy through the detection of a
gravitational signal coming from the merger of two astrophysical
black holes. Such signal was strong enough to permit the observation
of the ringdown phase characterized by the so-called quasinormal modes
(QNMs), which carry information of the structure of the spacetime
itself. In addition, the study of QNMs spectra can bring a better
understanding of the stability of a given black hole solution
\cite{Regge:1957td,Berti:2009kk,Kokkotas:1999bd,Nollert:1999ji,Konoplya:2008rq}.
Moreover, this question can be addressed
through the scattering of a scalar field in the fixed black hole background \cite{Abdalla:2005hu,Wang:2000dt,Wang:2004bv,Abdalla:2008te,Abdalla:2019irr,CuadrosMelgar:2011up}, which can be understood as a probe field to test the (in)stability of the black hole metric.

The QNMs and its spectrum are characterized, under appropriate boundary conditions, by a set of complex frequencies and encode the linear response of the black hole geometry to an external probe field with different spin weights. The time evolution of such probe fields is divided in three main stages: the initial burst in a short interval depending on the initial conditions, followed by the damping oscillation given by the QNMs and, at late-times, a power-law or exponential tails. 

Another interesting subject that black holes bring is their
thermodynamics. The similarity between classical thermodynamics and
the laws governing the mechanics of black holes was well established
by Bekenstein and Hawking~\cite{bek,haw} through the identification of
black hole surface gravity and event horizon area with the temperature and
entropy of a thermodynamical system, respectively. This fact led to the well known
Bekenstein-Hawking formula, 
\be 
\label{bekhaw}
S_{BH} = \frac{Area}{4}\,,
\ee
expressed in geometrical units. Based on this novel theory Bekenstein
proposed the existence of an upper bound on the entropy of any system
of energy $E$ and dimension $R$ given by $S\leq 2\pi
ER$~\cite{PhysRevD.23.287}. This equation is a consequence of the
validity of the generalized second law (GSL) of black hole
thermodynamics. Furthermore, in an effort to include quantum aspects
in the gravitational theory describing a black hole, 't
Hooft~\cite{tHooft:1984kcu} proposed 
a semi-classical method to compute the corrections to the classical
entropy formula (\ref{bekhaw}). This technique
known as the brickwall method consists in considering a thermal bath
of scalar fields living outside the event horizon. The quantization of
these fields via statistical mechanics partition function leads to
quantum corrections to the black hole entropy. By carrying out this
calculation on a Schwarzschild black hole 't Hooft showed that the
first correction is proportional to the area, as expected, having a
coefficient dependent on the proper distance from the horizon to the
wall. Later calculations in other solutions showed that this first
correction is the same in 4-dimensional geometries.

In this work we are interested in a solution of Einstein equations discovered by
Kiselev~\cite{kiselev}, which describes a spherically symmetric black
hole surrounded by an anisotropic
fluid~\cite{Visser:2019brz,Boonserm:2019phw}. This constitutes a line-element derived from the solutions studied in~\cite{Visser:1992qh}, the so-called dirty black holes. Studies on its stability
\cite{deOliveira:2018weu,Chen:2005qh,Guo:2013mna,Zhang:2006hh,Varghese:2014xaa} and some aspects of its thermodynamical behavior have been
implemented in the last
years~\cite{thomas,ghaderi,rodrigue,toledo,saheb}. However, a 
detailed description of the causal 
structure of the spacetime, the late-time behavior of the scalar QNMs,
and other aspects related to corrections to the entropy and thermodynamical stability are absent in the literature. 

The paper is organized as follows, Section \ref{sec2} presents the
metric describing the family of black holes surrounded by anisotropic
fluid and its main features. In Section \ref{sec3} we present the
causal structure of this spacetime. Also, the perturbative dynamics
due to probe scalar field evolution is formulated and the QNMs
spectrum and late-time tails are computed. Section \ref{sec4} brings a
study of some aspects of black hole thermodynamics including
Bekenstein entropy bound, semiclassical corrections to entropy through
t'Hooft brick wall method, and thermodynamical stability tested using
specific heat and Hessian matrix criteria. Finally, in Section
\ref{sec5} some final comments are given.

\section{Black Hole Solutions}\label{sec2}

We are interested in a kind of dirty black hole whose
line-element can be written as
\be
\label{metric}
ds^2 = -f(r)\,dt^2 + \frac{dr^2}{f(r)} + r^2 d \Omega^2 \,,
\ee 
where $d \Omega^2$ represents the metric of the 2-sphere and $f(r)$ is given
by~\cite{kiselev}
\be
\label{mcoeff}
f(r)=1-\frac{2M}{r} + \frac{Q^2}{r^2}-\frac{c}{r^{3\omega_f+1}},
\ee 
being $M$ the black hole mass, $Q$ its electric charge, $c=r_q^{3\omega_f +1}$ a constant ($r_q$ is a dimensional normalization constant), and $\omega_f$ a
parameter that characterizes an anisotropic fluid surrounding the black
hole, obeying the equation of state $p_f=\omega_f \rho_f$. 
Concerning this line-element there are four very special cases
depending on the value of the state parameter. 
The value $\omega_f=-1$ corresponds to a
Reissner-Nordstr\"om-(Anti)-de Sitter black hole where $3c$ plays the
role of the cosmological constant. 
When $\omega_f=-1/3$, we have a topological Reissner-Nordstr\"om black
hole. 
If $\omega_f=0$, the solution describes a Schwarzschild spacetime with
shifted mass. 
And for $\omega_f=1/3$ the metric corresponds to a Reissner-Nordstr\"om black hole whenever $c<Q^2$.

For the line-element (\ref{metric}) with the metric coefficient
(\ref{mcoeff}) for all possible values of fluid state parameter a
relative pressure anisotropy of the spacetime is defined by ~\cite{Visser:2019brz,Boonserm:2019phw}
\be
\label{eq3}
\Delta = \frac{p_r-p_t}{(p_r+2p_t)/3}= -\frac{3}{2}\left[\frac{4Q^2 - c \omega_f (1+\omega_f )r^{1-3\omega_f} }{Q^2 - c \omega_f^2 r^{1-3\omega_f} }\right]\,,
\ee 
where $p_r$ and $p_t$ represent the total energy-momentum tensor components
$T_{11}$ and $T_{22}=T_{33}$, respectively. This non-zero anisotropy 
labels a non-quintessential fluid, different from what was stated in
the first work which presented such a metric~\cite{kiselev}.

Furthermore, we can reinterpret the energy-momentum tensor of the
solution as a sum of anisotropic fluids with different state
parameters instead of considering a black hole surrounded by just one fluid component. By writing 
\be
\label{gtt}
g_{tt}= -\sum_{n}\frac{c_n}{r^{f_n}}\,,
\ee
with $c_n$ and $f_n$ being constants, the energy-momentum tensor is linear in each 'charge' $n$, i.e., $T= T^{f_1}+T^{f_2}+T^{f_3}+\cdots$. In such case by the proper choice of $c_n$'s and $f_n$'s we can easily have the charged black hole surrounded by a fluid as represented previously, meaning that the traditional components of charge and mass can be seen as fluid charges in the Kiselev picture~\cite{kiselev}.

Now the null energy condition imposes severe restrictions on the state
parameter of the fluid $\omega_f$. By taking the condition of validity
of the null energy statement~\cite{Visser:2019brz,Boonserm:2019phw} we
have that the density gradient of the fluid is
\be
\label{cc1}
\rho'=\left(\frac{m'}{4\pi r^2}\right)' \leq 0\,,
\ee
where $m'$ represents the derivative of the position-dependent mass
function $m(r)$ defined as~\cite{Boonserm:2019phw}
\be
\label{massfluid}
2\,m(r) = \sum_{i=0} ^N K_i \, r^{-3\omega_i}\,,
\ee
with $K_i$ and $\omega_i$ being general coefficients and exponents of a
Puiseux series. 
In our case we obtain
\be
\label{cc2}
\rho'=\frac{1}{8\pi r^4}\left[ -\frac{4Q^2}{r}+\frac{9c \omega_f (\omega_f + 1)}{r^{3\omega_f}}\right]\,.
\ee
Thus, the energy condition is preserved whenever $-1 \leq \omega_f
\leq 0$, and violated otherwise. For this reason in this work we will study dynamical and thermodynamical aspects of the geometry within the range of validity of such condition.

In the next section we are going to characterize the causal structure
of the family of solutions represented by the line-element
(\ref{metric}) establishing the nature of the singularity and the
horizons. In addition, we will check the late-time behavior of scalar
QNMs in that geometry.

\section{Causal Structure and Probe Scalar Field Evolution}\label{sec3}

We are going to describe the causal structure for two different
representative black hole solutions of the metric (\ref{metric}). 
We start by considering the behavior of the Kretschmann invariant
given by
\begin{equation}\label{curvature_invariant}
R_{abcd}R^{abcd}= \frac{48M^2}{r^6} - \frac{96MQ^2}{r^7} + \frac{56Q^4}{r^8} + \frac{ c^2p_1}{r^{2(2+\sigma)}} + \frac{8 c M  p_2}{r^{(5+\sigma)}} -  \frac{4 c Q^2 p_3}{r^{(6+\sigma)}},
\end{equation}
where we have defined $\sigma=3w_f +1$,
$p_1=\sigma^4+2\sigma^3+5\sigma^2 + 4 $, $p_2=\sigma^2+3\sigma+2$ and
$p_3=3\sigma^2+7\sigma+2$. In the cases when $w_f\leq 0$ we have
$\sigma\leq 1$, so the Kretschmann invariant always diverges at $r=0$ and is well behaved at the horizons and, thus, the line-element (\ref{metric}) has a physical singularity at the origin $r=0$. In what follows, we are going to show that for two specific cases $\omega_f=-1/2$ and $\omega_f= -2/3$ with $M>Q$ there is a range of parameters that represents a black hole with cosmological-like horizon $r_c$, an event horizon $r_{+}$, and Cauchy inner horizon $r=r_{-}$ covering the time-like singularity at $r=0$. Such causal structure is very similar to the Reissner-Nordstr\"om-de Sitter black hole, except in the region beyond the cosmological-like horizon $r>r_c$, where the spatial infinity ($r\rightarrow \infty$) is light-like.
\subsection{Black hole solution with $w_{f}=-1/2$}
Considering the line-element (\ref{metric}) with $w_f=-1/2$ and the redefinition of the radial coordinate $r=z^2$ we have
\begin{equation}\label{wf_1_2}
ds^2= - \frac{c}{z^4}H(z)dt^2 +\frac{4z^6}{c}H(z)^{-1}dz^2 + z^4 d\Omega^{2},
\end{equation}
where the function $H(z)$ is given in terms of three real roots $z_c>z_{+}>z_{-}$ denoting, respectively, the cosmological-like, event, and Cauchy horizons, and two real negative roots $(z_{1},z_{2})$. Thus, 
\begin{equation}\label{h_1_2}
 H(z)=-(z-z_{c})(z-z_{+})(z-z_{-})(z+z_1)(z+z_2),
\end{equation}
yields a tortoise coordinate given by
\begin{equation}\label{tortoise_1_2}
 z_{*}=-\frac{2}{c}z-\alpha_{c}\log{|z-z_c|}+\alpha_{+}\log{|z-z_+|} - \alpha_{-}\log{|z-z_-|} + \alpha_{1}\log{|z+z_1|}-\alpha_{2}\log{|z+z_2|},
\end{equation}
which defines the usual double null system, $U=t-z_{*}$ and $V=t+z_{*}$. Here the constants $(\alpha_c, \alpha_+, \alpha_{-},\alpha_1,\alpha_2)$ are all positive definite and are given in terms of the horizons 
\begin{equation}\label{alphas}
\alpha_{i} = \frac{2z_{i}^{5}}{c}\prod_{i\neq j}\frac{1}{|z_{i}-z_{j}|},
\end{equation}
where the indices $i$ and $j$ denote the horizons $(z_{c}, z_+, z_{-}, z_{1},z_{2})$. 

We perform a detailed examination of the behavior of the black hole
solution in the vicinity of each horizon in order to obtain the
Kruskal-Szekeres extension to end up with the Penrose-Carter diagram
of the entire manifold. 

Near the cosmological-like horizon $z=z_c$, the Kruskal-Szekeres coordinates $U_{c}$ and $V_{c}$ obey the following relation
\begin{equation}\label{uv_1_2_rc}
U_c V_c = \pm e^{(2/c\alpha_c)z}|z-z_c|\left(\frac{|z-z_-|^{\alpha_-}}{|z-z_{+}|^{\alpha_+}}\frac{|z+z_2|^{\alpha_2}}{|z+z_1|^{\alpha_1}}\right)^{1/\alpha_c},
\end{equation}
where the plus sign denotes the region $z>z_c$ and the negative sign
corresponds to the region $z<z_c$. Similarly, near the event horizon
$z_{+}$ we have
\begin{equation}\label{uv_1_2_rh}
U_+ V_+ = \mp e^{-(2/c\alpha_+)z}|z-z_+|\left(\frac{1}{|z-z_c|^{\alpha_c}}\frac{1}{|z-z_-|^{\alpha_-}}\frac{|z+z_1|^{\alpha_1}}{|z+z_2|^{\alpha_2}}\right)^{1/\alpha_+},
\end{equation}
where the upper sign refers to $z>z_+$ and the lower sign refers to $z<z_+$. Finally, for the region near the Cauchy horizon $z\approx z_-$, we have
\begin{equation}\label{uv_1_2_rin}
U_{-}V_{-}=\pm e^{2/c\alpha_-}|z-z_-|\left(\frac{|z-z_c|^{\alpha_c}}{|z-z_+|^{\alpha_+}}\frac{|z+z_2|^{\alpha_2}}{|z+z_1|^{\alpha_1}}\right)^{1/\alpha_-}.
\end{equation}

\begin{figure}[!ht]
\begin{center}
\includegraphics[scale=0.7]{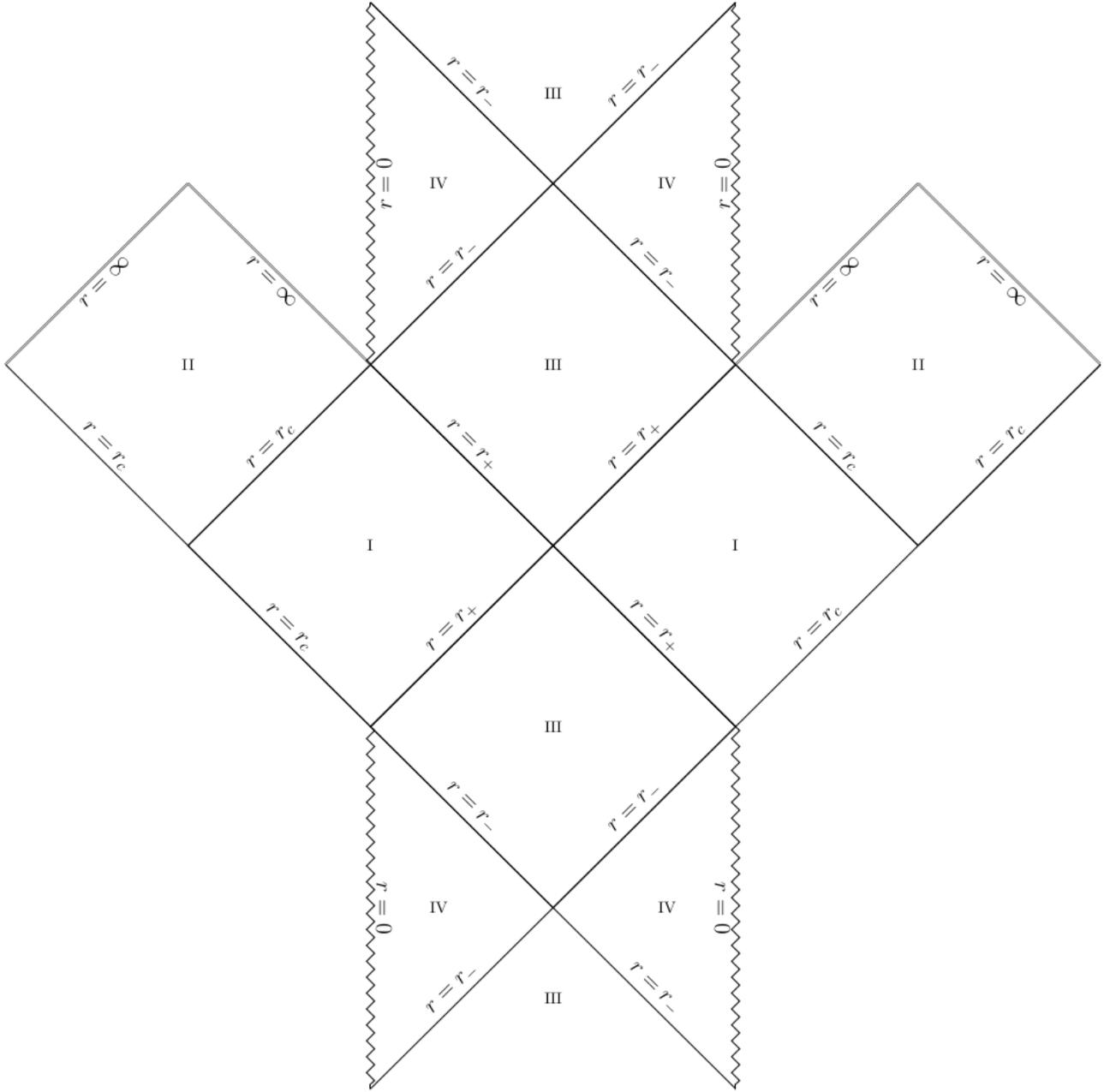}
\end{center}
\caption{Penrose-Carter diagram for the four-dimensional black hole with $w_{f}=-1/2$ and $w_f=-2/3$}
\label{penrose_1_2}
\end{figure}

Introducing the Penrose coordinates $T=\frac{1}{2}(\tilde{U}+\tilde{V})$ and $R=\frac{1}{2}(\tilde{U}-\tilde{V})$ in each region covered by the relations (\ref{uv_1_2_rc} - \ref{uv_1_2_rin}) with $\tilde{U}=\arctan(U)$ and $\tilde{V}=\arctan(V)$, we  compactified the coordinates. Furthermore, combining different overlaping coordinate patches it is possible to extend the metric through each horizon, thus, constructing the conformal diagram for the entire spacetime (\ref{wf_1_2}) in Fig. (\ref{penrose_1_2}). Such diagram shows a causal structure very similar to that of a Reissner-Nordstr\"om-de Sitter black hole ~\cite{Griffiths:2009dfa,Laue:1977zz}. We observe an infinite sequence of structures featuring two outer horizons (event and cosmological-like), an inner Cauchy horizon, and a time-like singularity at the origin $z=0$. However, the spatial infinity ($z\rightarrow \infty$) in the black hole solution with $w_f=-1/2$ displays a light-like structure, which is different from the Reissner-Nordstr\"om-de Sitter case, where the spatial infinity is space-like (see Fig.2 in \cite{Laue:1977zz}).

For an observer in region I crossing the event horizon and entering region III, we observe that the coordinate $z$ is now time-like and the subsequent motion occurs with $z$ decreasing. However, after the observer crosses the Cauchy horizon, the coordinate $z$ becomes space-like again, so it is possible for this observer to avoid the time-like singularity at $z=0$ and emerge in another copy of region III.

The maximally extended black hole with $w_f=-2/3$, and the conformal diagram is the same as in the case $w_f=-1/2$, and can be obtained by performing the same steps as discussed here. The detailed calculation of the extension is given in the Appendix \ref{apendiceB}.

\subsection{Klein-Gordon equation}

For a black hole spacetime as represented in Fig.\ref{penrose_1_2} the physical universe lies in region I, where we choose to integrate a scalar field that do not change the geometry. 

In this domain the integration of the Klein-Gordon equation, $\Box
\Phi =0$, will be affected by the definition of a tortoise coordinate
system, $dx=f^{-1}dr$, (now in terms of $r$) used to fix the field
propagation as ingoing plane waves crossing through the boundaries of $x$. In terms of this system the field equation turns to the usual simple form
\be
\label{eq4}
\left( \frac{\partial^2 }{\partial x^2} - \frac{\partial^2}{\partial t^2 } -V(r) \right) \Psi = 0, 
\ee
where $\Psi$ represents the radial-temporal part of the Klein-Gordon field written as 
\be
\label{eq5}
\Phi = \sum_{l,m} Y_l^m(\theta , \phi ) \frac{\Psi (r,t)}{r},
\ee
and $V(r)$ plays the role of a potential for the scattered scalar waves given by
\be
\nonumber
V(r) = f(r) \left[ \frac{\partial_r f(r)}{r} + \frac{l(l+1)}{r^2}  \right] = \hspace{5.0cm} \\
\label{eq6}
\left( 1-\frac{2M}{r} + \frac{Q^2}{r^2}-\frac{c}{r^{3\omega_f+1}}\right) \left[ \frac{2M}{r^3} - \frac{2Q^2}{r^4} + \frac{c (3\omega_f+1)}{r^{3\omega_f+3}} + \frac{l(l+1)}{r^2} \right]\,.
\ee
 
In de Sitter spacetimes the tortoise coordinate places the cosmological horizon $r\rightarrow r_c$ at the point $x=\infty$ and the event horizon of the black hole $r \rightarrow r_+$ at $x=-\infty$. This is also the case for dirty black holes with an anisotropic fluid as discussed in this paper. As a consequence, when using the above wave equation we will restrict the integration to the region $-\infty < x < \infty$ in block I of Penrose diagram.

When studying the evolution of fields in fixed geometries, Eq.(\ref{eq4}) establishes a master equation and for different fields (or spherical geometries) the proper $V(r)$ must be taken.

The numerical integration in double null-coordinates for the calculus of the quasinormal modes is a well-establish method, which in general does not depend on the initial conditions. Except for the “initial burst” of evolution, the quasinormal ringing phase that follows and the late-time behavior depend only on the geometry parameters. In terms of the null coordinate system $u$ x $v$,
\be
\nonumber
2dv & = & dt + dx \\
\label{eq7}
2du & = & dt - dx,
\ee
the Klein-Gordon equation takes the form
\be
\label{eq8}
\left[ 4 \frac{\partial^2}{\partial u \partial v} +V(r) \right] \Psi = 0,
\ee
or, written as a discrete equation,
\be
\label{eq9}
\Psi_N= \Psi_W + \Psi_E - \Psi_S - \frac{h^2}{8} V_S \left[ \Psi_W + \Psi_E \right]
\ee
The boundary conditions in such system can be put in the form 
\be
\label{r2}
 \Psi|_{fixed \hspace{0.1cm} v} =  constant, \hspace{0.4cm} \Psi|_{fixed \hspace{0.1cm} u} =  Gaussian \hspace{0.2cm} package,
\ee
although discussions on the preservation of polar and radial symmetry
(for the gravitational field) have presented Neumann boundary
condition as the appropriate one.

After obtaining the field profile in time domain we can employ the
Prony method~\cite{Konoplya:2011qq} to acquire the quasinormal
frequencies or, in the case of non-oscillatory profiles, linear
regression. We will also use the WKB6 method~\cite{Konoplya:2003ii,Schutz:1985zz,Iyer:1986np} as a matter of comparison.

\subsection{Late-time behavior and quasinormal modes}

\begin{figure}[!ht]
\begin{center}
\includegraphics[width=8.1cm, height=5.5cm]{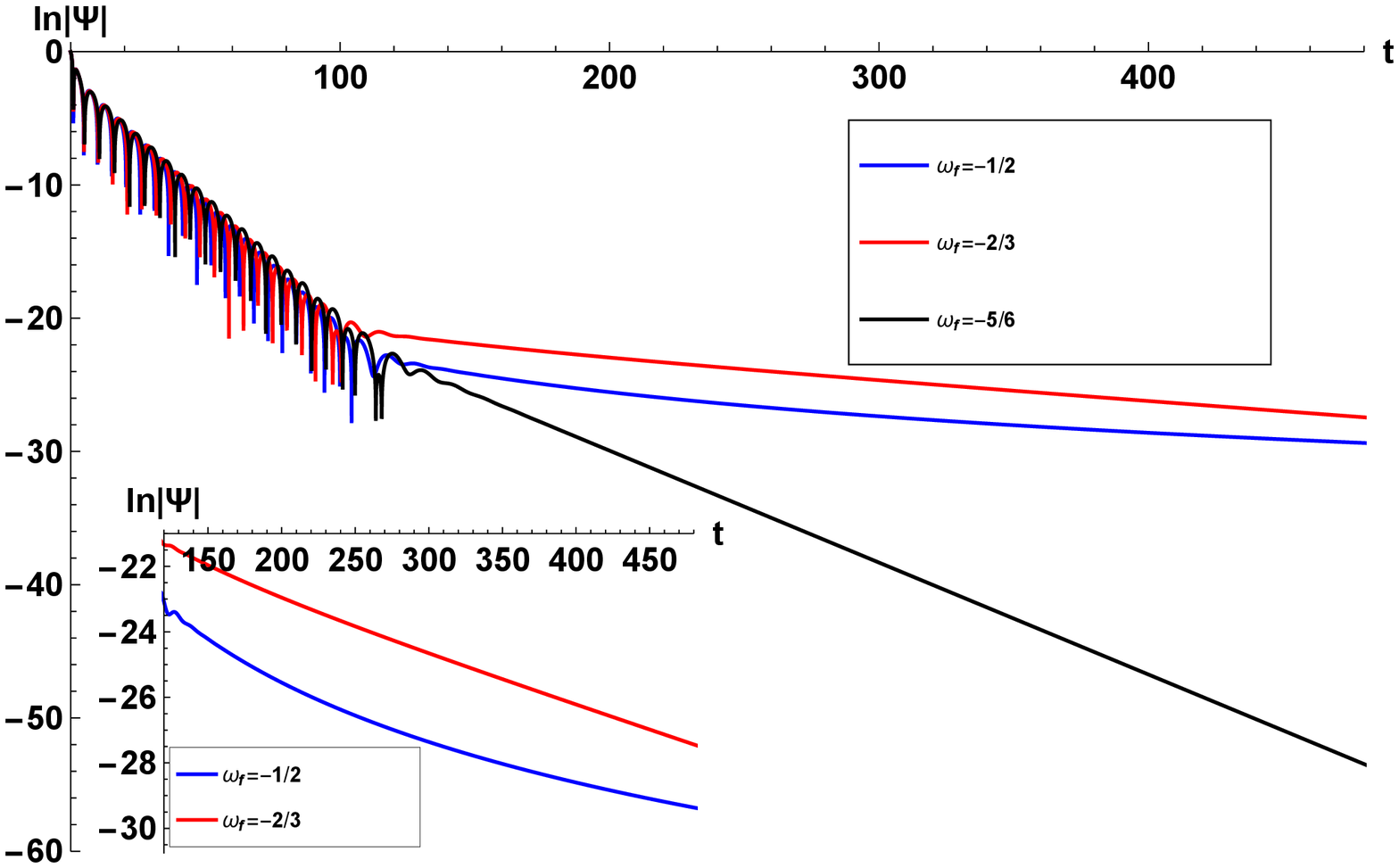}
\includegraphics[width=8.1cm, height=5.5cm]{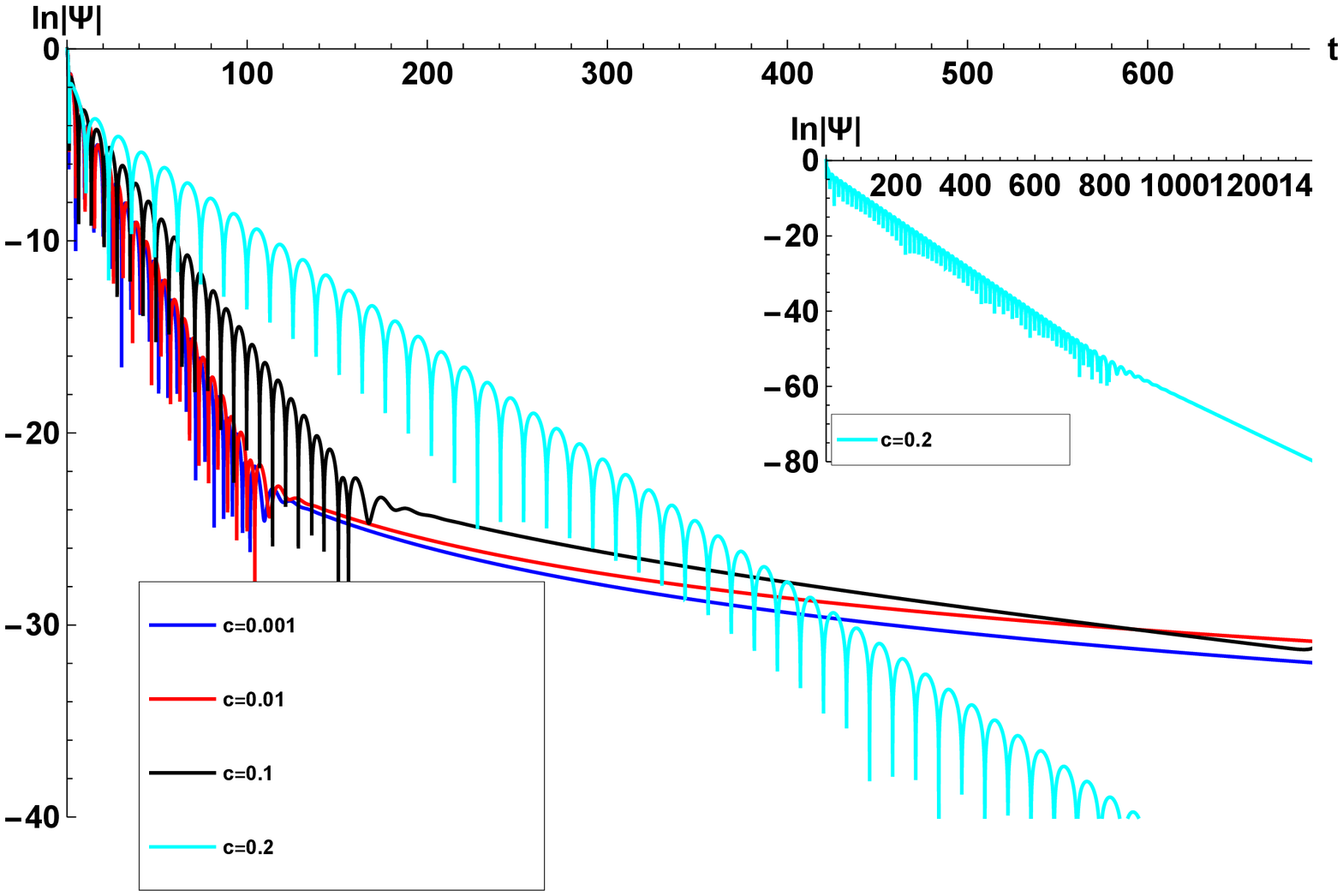} 
\end{center}
\caption{{\small Late-time behavior of propagating scalar field in dirty black holes, profiles of exponential decay or power-law tail. Left panel: $M=2Q=l=10c=1$; right panel: $M=2Q=l=-2\omega_f=1$.}}
\label{fig1}
\end{figure}

The late-time evolution of the probe scalar field brings two distinct
behaviors depending on the fluid parameters $c$ and $\omega_f$. In
Fig.\ref{fig1} we can see different field profiles evolved from a similar initial burst as defined above. Depending on the fluid charge parameters we have an exponential decay or a power-law tail dominating the final stage of evolution. The fact comes surprisingly as a combination of two distinct behaviors already found in black holes with/without cosmological constant, being such final stage an exponential decay/power-law tail, respectively, for the Reissner-Nordstr\"om case. 

In the left panel the exponential decay mode comes for the highest
frequencies $|\omega_f|$ and the power-law tail happens when
$\omega_f=-1/2$. The dominant purely imaginary quasinormal modes
(smallest imaginary part) also present in the de Sitter black holes
spacetimes~\cite{Du:2004jt} are a family of modes connected to the presence of the cosmological constant~\cite{Cardoso:2017soq} (or, in our case, to the anisotropic fluid density). For small enough values of the fluid state parameter $\omega_f$ and density $c$ though, the dominant term between horizons is that of the Schwarzschild potential generating the well-known power-law  behavior~\cite{Kokkotas:1999bd,Gundlach:1993tp,PhysRevD.5.2419}. Such role is associated to the weak decay of the potential for high values of $r$ and may come as a {\it general} result of the integral around the negative imaginary $\omega$ axis. The same qualitative characteristic can be seen on the right panel of Fig.\ref{fig1}. Whenever $c >~ 0.1$, the fluid term is dominant and a purely imaginary quasinormal mode overcomes the power-law tail behavior.

A region of frontier in the parameters happens approximately at
$-0.65\lesssim \omega_f \lesssim -0.5$, which is represented by an
exponential decaying final stage if $\omega_f < -0.65$ and a power-law
tail if $\omega_f > -0.5$. Between both values the dominant final
stage of the field in late-times alternates between these two
profiles, a feature we discuss in what follows.
\begin{table}[htbp]
  \centering
 \caption{The quasinormal modes of the RN black hole with anisotropic fluid. The parameters of the geometry and scalar field read $M=2Q=l=1$.}
    \begin{tabular}{ccccccc}
    \hline
 & \multicolumn{2}{c}{$\omega_f=-1/2$} \hspace{1.0cm} &   \multicolumn{2}{c}{$\omega_f=-2/3$} \hspace{1.0cm}& \multicolumn{2}{c}{$\omega_f=-5/6$} \\
	\hline
    $c/c_{max}$  & {$\Re(\omega)$} & {$-\Im(\omega)$} &  {$\Re(\omega)$} & {$-\Im(\omega)$} &  {$\Re(\omega)$} & {$-\Im(\omega)$}  \\
\hline \hline
0.000001&	0.306577	&	0.098825	&	0.306577	&	0.098825	&	0.306577	&	0.098825 \\
0.001	&	0.306339	&	0.098731	&	0.306381	&	0.098756	&	0.306393	&	0.098774 \\
0.01	&	0.304192	&	0.097883	&	0.304614	&	0.098133	&	0.304735	&	0.098313 \\
0.1	&	0.282652	&	0.089421	&	0.286716	&	0.091821	&	0.287854	&	0.093546 \\
0.2	&	0.258577	&	0.080084	&	0.266302	&	0.084612	&	0.268423	&	0.087876 \\
0.3	&	0.234234	&	0.070970	&	0.245124	&	0.077112	&	0.248194	&	0.081799 \\
0.4	&	0.209281	&	0.061993	&	0.223141	&	0.069385	&	0.226918	&	0.075302 \\
0.5	&	0.183879	&	0.052833	&	0.200051	&	0.061351	&	0.204263	&	0.068156 \\
0.6	&	0.157742	&	0.043015	&	0.175489	&	0.052864	&	0.180136	&	0.060159 \\ 
0.7	&	0.130354	&	0.033828	&	0.148884	&	0.044002	&	0.153675	&	0.050825 \\
0.8	&	0.100866	&	0.024957	&	0.118713	&	0.034209	&	0.122685	&	0.042393 \\
0.9	&	0.067115	&	0.016471	&	0.081417	&	0.024291	&	0.085918	&	0.027818 \\
0.99	&	0.019522	&	0.003853	&	0.020814	&	0.011965	&	0.024821	&	0.012187 \\
   \hline
    \end{tabular}
  \label{t1}
\end{table}

A second element present in the scalar field evolution of the above figures is the quasinormal modes, damped oscillations that arrive given the presence of a black hole potential barrier such as (\ref{eq6}).
In table \ref{t1} we list the fundamental modes for different values
of fluid density. As expected, the influence of the fluid in the
scalar field QNMs is very mild when its density is small (not
detectable, e.g. for $c \sim 10^{-6}$), no matter what the state
parameter is. As $c$ increases, the differences coming from several
state parameters of the fluid increase as well. We can see that the
quality factor, $\mathbb{Q} =  \frac{\Re(\omega)}{-\Im(\omega)}$, decreases as
we increase $|\omega_f|$. In fact, in a spacetime with an anisotropic
fluid the scalar field oscillates better compared to a spacetime with
cosmological constant: e. g. when $M=2Q=l=10c/3c_{max}=1$, we have $\mathbb{Q}=3.30, 3.18, 3.03$ and $2.88$ for $\omega_f=-1/2,-2/3,-5/6$ and $-1$, respectively. 

The results in the Table \ref{t1} were double checked with the WKB6
method~\cite{Konoplya:2003ii}. The convergence of both calculations is
as good as $0.1\%$ for $c/c_{max} \lesssim 0.5$, where $c_{max}$ represents the maximum value of fluid density to which 3 horizons arise. Whenever the fluid density is high, higher divergences are found. This comes as no surprise as long as the WKB6 has a poor convergence for near extremal black holes. 

For a large range of parameters we investigate the transitional behavior of the scalar field at late-times. Testing for the linear correlation of two different profiles written as
\be
\label{ap3a}
\Psi|_{late \hspace{0.1cm} times} & \rightarrow & t^{-a} \,,\\
\Psi|_{late \hspace{0.1cm} times} & \rightarrow & e^{-\alpha t}\,,
\ee
we perform calculations for different state parameters going from $\omega_f = -0.5$ to $\omega_f = -1$. The results are given in Table \ref{tap1}. 
\begin{table}[htbp]
  \centering
 \caption{Quasinormal modes and late-time behavior of the RN black hole with anisotropic fluid in a geometry with $M=2Q=100c=1$. The scalar field angular momentum reads $l=1$.}
    \begin{tabular}{cccccccc}
    \hline
$-\omega_f$ & $a$ & $R^2$ & $\alpha$ & $R^2$ & $\Re(\omega)$ & $-\Im(\omega)$ \\ 
\hline \hline
0.52	&	4.88671		&	1.00000	&	0.01551	&	0.98920	&	0.29729	&	0.095191 \\
0.55	&	2.29142		&	0.99582	&	0.00611	&	0.99981	&	0.29629	&	0.094800 \\
0.58	&	2.87836		&	0.99906	&	0.00917	&	0.99459	&	0.29518	&	0.094512 \\
0.61	&	3.18710		&	0.99933	&	0.01014	&	0.99398	&	0.29395	&	0.094148 \\
0.64	&	3.79916		&	0.99723	&	0.01213	&	0.99741	&	0.29260	&	0.093758 \\
0.67	&	6.07792		&	0.99753	&	0.01618	&	0.99997	&	0.29108	&	0.093323 \\
0.7	&	9.01819		&	0.99703	&	0.02402	&	1.00000	&	0.28941	&	0.092925 \\
0.73	&	12.82665	&	0.99698	&	0.03416	&	1.00000	&	0.28755	&	0.092521 \\
0.76	&	17.40289	&	0.99697	&	0.04635	&	1.00000	&	0.28550	&	0.092112 \\
0.79	&	26.61588	&	0.99697	&	0.07089	&	1.00000	&	0.28159	&	0.091429 \\
0.82	&	28.67909	&	0.99697	&	0.07639	&	1.00000	&	0.28073	&	0.091293 \\
0.85	&	35.26838	&	0.99697	&	0.09394	&	1.00000	&	0.27795	&	0.090884 \\
0.88	&	42.40377	&	0.99697	&	0.11294	&	1.00000	&	0.27488	&	0.090478 \\
0.91	&	50.01264	&	0.99697	&	0.13321	&	1.00000	&	0.27150	&	0.090076 \\
0.94	&	72.31642	&	0.99753	&	0.15068	&	0.99759	&	0.26776	&	0.089683 \\
   \hline
    \end{tabular}
  \label{tap1}
\end{table}
Observing the high values of linear correlation we state that both behaviors (exponential decay and power-law) are present in the final stage of the field evolution being one of them dominant. 

We can see a small variation in the linear coefficients of the power-law series for $-0.65 \lesssim \omega_f \lesssim -0.5$ and an explosion after that, softening its behavior in the field composition $\Psi|_{late \hspace{0.1cm} times} \rightarrow C_1t^{-a} + C_2e^{-\alpha t}$. This makes the presence of this term subdominant in relation to the exponential decay series, which is prevalent for $\omega_f \gtrsim -2/3$. 

This comes as an interesting result not stated until now in the available literature, e. g. for RNdS geometries, the presence of a power-law tail term subdominant to the imaginary quasinormal mode in late-times in such spacetimes. 

In the last two columns of the Table \ref{tap1} we can see the
quasinormal modes frequencies for a variety of $\omega_f$. The
frequencies were obtained via Prony method with the same field
profiles used in the late-time test. Again they were checked with WKB6 method with very good agreement in the results (maximum deviation of $0.1\%$).

\section{Thermodynamics}\label{sec4}

In this section we are going to discuss some thermodynamical aspects
of the dirty black holes under consideration. 

First of all, we can rewrite the metric coefficient (\ref{mcoeff}) in
terms of the event horizon as
\begin{equation}\label{met}
f(r) = \frac{r-r_+}{r} - \frac{Q^2}{r^2}\frac{(r-r_+)}{r_+} +
\frac{c}{r^{3\omega_f+1}} \frac{(r^{3\omega_f}-r_+ ^{3\omega_f})}{r_+ ^{3\omega_f}}\,.
\end{equation}
In addition, using the metric (\ref{metric}) we can write the surface gravity as
\begin{equation}\label{kappa}
\kappa = \frac{1}{2} f'(r)\rvert_{r=r_+} = \frac{1}{2}
\left[\frac{2M}{r_+ ^2}-\frac{2Q^2}{r_+ ^3}+(3\omega_f +1)\frac{c}{r_+ ^{3\omega_f+2}}\right]\,.
\end{equation}
Both expressions will be useful in our next calculations.

\subsection{Entropy Bound}

Let us consider a particle in equatorial motion near a black hole. The
constants of motion are given by
\begin{eqnarray}\label{com}
E &=& \pi_t = g_{tt} \dot t \nonumber \\
J &=& -\pi_\phi = -g_{\phi\phi} \dot\phi \,,
\end{eqnarray}
corresponding to the energy and angular momentum of the
particle, respectively. Since the energy conservation for a particle of mass $m$
implies $-m^2=\pi_\mu \pi^\mu$, using the metric (\ref{metric})
together with the metric coefficient (\ref{mcoeff}) we can obtain a
quadratic equation for the conserved energy $E$ of the particle,
\begin{equation}\label{quadeq}
E^2 - \frac{fJ^2}{r^2} - m^2 f = 0 \,,
\end{equation}
whose solution becomes
\begin{equation}\label{solE}
E = \sqrt{m^2 f + \frac{fJ^2}{r^2}} \,.
\end{equation}
As the particle is approaching the black hole gradually, this process
must stop when the proper distance from the body's center of mass to
the black hole horizon equals the body's radius $R$,
\begin{equation}\label{R}
\int_{r_+} ^{r_+ +\delta(R)} \sqrt{g_{rr}} \,dr = R \,,
\end{equation}
where $r_+ + \delta(R)$ represents the point of capture of the
particle by the black hole.  At this point the energy of the particle (\ref{solE}) can
be evaluated and minimized with respect to the angular momentum of the
particle. This results in $J_{min}=0$, such that
\begin{equation}\label{emin}
E_{min} = \sqrt{f(r_++\delta)}\, m \,.
\end{equation}

In order to perform the integral (\ref{R}), express $\delta$ in terms
of $R$, and evaluate Eq.(\ref{emin}), we considered 3 cases, $\omega_f
= -1/2,\; -2/3,\; -5/6$. To first order in $\delta$ the proper
distance integral yields,
\begin{eqnarray}\label{del}
\delta = \left\{
\begin{array}{ll}
\frac{(2r_+ ^2 r_q -2Q^2r_q -3r_+ ^2\sqrt{r_+ r_q})R^2}{8r_+ ^3 r_q}
\,, \quad &\hbox{for} \quad \omega_f = -1/2 \\
\frac{(r_+ ^2 r_q -Q^2 r_q -2r_+ ^3)R^2}{4r_+ ^3 r_q} \,, \quad
&\hbox{for} \quad \omega_f = -2/3 \\
\frac{(2r_+ ^2 r_q ^2 -2Q^2 r_q ^2 - 5r_+ ^3 \sqrt{r_+ r_q})R^2}{8r_+ ^3
  r_q ^2} \,, \quad &\hbox{for} \quad \omega_f =-5/6 
\end{array} \right.
\end{eqnarray}
From the first law of thermodynamics we have that 
\begin{equation}\label{1law}
dM = \frac{\kappa}{2}\, dA_r \,,
\end{equation}
being $A_r$ the rationalized event horizon area $A/4\pi$ and
$dM=E_{min}$, the change 
in the black hole mass due to the assimilation of the particle. Using
Eqs.(\ref{kappa}), (\ref{emin}), and (\ref{del}) we obtain 
\begin{equation}
dA_r = 2mR\,,
\end{equation}
in the three cases considered here. Now assuming the validity of the Generalized
Second Law (GSL), $S_{BH} (M+dM) \geq S_{BH}(M) + S$, we derive an
upper bound to the entropy $S$ of an arbitrary system of proper energy
$E$, 
\begin{equation}
S \leq 2\pi mR \,.
\end{equation}
This result is independent of the black hole parameters and perfectly
agrees with the universal bound found by Bekenstein~\cite{PhysRevD.23.287}. 

\subsection{Semiclassical corrections to black hole entropy}

Following 't Hooft's brickwall method~\cite{tHooft:1984kcu} we consider a
thermal bath of scalar fields propagating just outside the horizon of a black
hole background given by Eqs.(\ref{metric}) and (\ref{mcoeff}). The
minimally coupled scalar field with mass $\mu$ satisfies Klein-Gordon equation,
\begin{equation}\label{kg}
\frac{1}{\sqrt{-g}} \partial_\mu (\sqrt{-g} g^{\mu\nu} \partial_\nu
\Phi) - \mu^2 \Phi = 0 \,.
\end{equation}
The idea is to quantize this field using the partition function of
statistical mechanics, whose leading contribution comes from the
classical solutions of the Euclidean action that yield the
Bekenstein-Hawking formula. This scalar field will produce quantum
corrections to the black hole entropy which can be calculated using the
brickwall method. 
The 't Hooft method consists in introducing an ultraviolet cut-off near
the event horizon such that $\Phi=0$ for $r \leq r_+ +\epsilon$. In
addition, in order to eliminate infrared divergences another cut-off
is introduced at a large distance from the horizon, $L \gg r_+$, where
$\Phi=0$ for $r\geq L$. By decomposing the scalar field as
\begin{equation}
\Phi (t,r,\theta, \phi) = e^{-iEt} R(r) Y(\theta, \phi) \,,
\end{equation}
the radial part of Eq.(\ref{kg}) turns into
\begin{equation}\label{radial}
R'' + \left( \frac{f'}{f} + \frac{2}{r} \right) R' + \frac{1}{f}
\left[ \frac{E^2}{f} - \frac{\ell(\ell +1)}{r^2} - \mu^2 \right] R = 0\,,
\end{equation}
where $\ell (\ell +1)$ is the variable separation constant.
Then, using a WKB approximation for $R(r) \sim e^{iS(r)}$ in
Eq.(\ref{radial}), where $S(r)$ is a rapidly varying phase, to leading
order only the contribution from the first derivative of $S$ is
important. This contribution represents the radial wave number
$K\equiv S'$, which can be obtained from the real part of
Eq.(\ref{radial}) as
\begin{equation}\label{radK}
K = \frac{1}{\sqrt{f}} \left[ \frac{E^2}{f} - \frac{\ell (\ell
    +1)}{r^2} - \mu^2 \right]^{1/2} \,.
\end{equation}
In terms of this quantity the number of radial modes $n_r$ is quantized semiclassically as,
\begin{equation}\label{cond}
\pi n_r = \int_{r_+ +\epsilon} ^L K(r,\ell,E) dr \,.
\end{equation}

Furthermore, the entropy of the system will be calculated from the
Helmholtz free energy $F$ of the thermal bath of scalar particles with
temperature $\beta^{-1}=\kappa/2\pi$, 
\begin{equation}\label{f1}
F = \frac{1}{\beta} \int d\ell \, (2\ell +1) \int dn_r \, \ln(1-e^{-\beta
  E}) = -\int d\ell \, (2\ell +1) \int \frac{n_r}{e^{\beta E}-1} dE \,,
\end{equation}
where we made an integration by parts in the last step. Using
Eqs.(\ref{radK}) and (\ref{cond}) and performing the integral in
$\ell$ we obtain
\begin{equation}\label{f2}
F = - \frac{2}{3\pi} \int \frac{dE}{e^{\beta E}-1} \int_{r_+
  +\epsilon} ^L dr \left[\frac{r^2}{\sqrt{f}} \left(\frac{E^2}{f}
    -\mu^2 \right)^{3/2}\right] \,.
\end{equation}
According to brickwall method we should study the contribution of this
integral near the horizon. Thus, using Eq.(\ref{met}) to write an
approximate expression of the metric near the horizon and performing
the integral in $E$ we get
\begin{equation}\label{f3}
F \approx -\frac{2\pi^3}{45\beta^4} \int_{1+\bar\epsilon} ^{\bar L} r_+ ^3
\left[ \left(1-\frac{Q^2}{r_+ ^2} \right) (y-1) + \frac{c}{r_+ ^{3\omega_f
      +1}} (y^{3\omega_f} -1)\right]^{-2} dy \,,
\end{equation}
where we rescaled some quantities as $y=r/r_+$, $\bar L=L/r_+$, and 
$\bar\epsilon = \epsilon /r_+$.
At this point it is convenient to consider different values of
$\omega_f$ separately. We should notice that the divergent
contribution of the integral to the Helmholtz energy comes from its
lower limit. Thus, the leading divergent term $F_\epsilon$ is given by
\begin{eqnarray}\label{Fe}
F_\epsilon = \left\{
\begin{array}{ll}
-\frac{8\pi^3 r_+ ^4}{45\beta^4 \epsilon} \left(2 - \frac{2Q^2}{r_+
    ^2} - 3\sqrt{\frac{r_+}{r_q}}\right)^{-2}
\,, \quad &\hbox{for} \quad \omega_f = -1/2 \\
-\frac{2\pi^3 r_+ ^4}{45\beta^4 \epsilon} \left(1-\frac{Q^2}{r_+ ^2} -
\frac{2r_+}{r_q}\right)^{-2}  \,, \quad
&\hbox{for} \quad \omega_f = -2/3 \\
-\frac{8\pi^3 r_+ ^4}{45\beta^4\epsilon} \left[2-\frac{2Q^2}{r_+ ^2}
  - 5 \left(\frac{r_+}{r_q}\right)^{3/2} \right]^{-2}  \,, \quad
&\hbox{for} \quad \omega_f =-5/6 
\end{array} \right.
\end{eqnarray}

The corresponding entropy $S_\epsilon = \beta^2 \frac{\partial F}{\partial
\beta}$, then, becomes
\begin{eqnarray}\label{Se}
S_\epsilon = \left\{
\begin{array}{ll}
\frac{32\pi^3 r_+ ^4}{45\beta^3 \epsilon} \left(2 - \frac{2Q^2}{r_+
    ^2} - 3\sqrt{\frac{r_+}{r_q}}\right)^{-2} \,, \quad &\hbox{for} \quad \omega_f = -1/2 \\
\frac{8\pi^3 r_+ ^4}{45\beta^3 \epsilon} \left(1-\frac{Q^2}{r_+ ^2} -
\frac{2r_+}{r_q}\right)^{-2} \,, \quad &\hbox{for} \quad \omega_f = -2/3 \\
\frac{32\pi^3 r_+ ^4}{45\beta^3 \epsilon} \left[2-\frac{2Q^2}{r_+ ^2}
  - 5 \left(\frac{r_+}{r_q}\right)^{3/2} \right]^{-2} \,, \quad &\hbox{for} \quad \omega_f = -5/6 
\end{array} \right.
\end{eqnarray}
We can express our results in terms of the proper thickness $\alpha$ defined as
\begin{equation}\label{thick}
\alpha = \int_{r_+} ^{r_+ +\epsilon} \sqrt{g_{rr}}\, dr \,.
\end{equation}
To first order this expression can give us a relation between $\epsilon$ and
$\alpha$ for the values of $\omega_f$ considered here,
\begin{eqnarray}\label{ea}
\epsilon \approx \left\{
\begin{array}{ll}
-\frac{\alpha^2}{8 r_+ ^3} \left(2Q^2 - 2r_+ ^2 + 3\frac{r_+
  ^{5/2}}{\sqrt{r_q}} \right)\,, \quad &\hbox{for} \quad \omega_f = -1/2 \\
-\frac{\alpha^2}{4r_+ ^3 r_q} \left(Q^2 r_q + 2r_+ ^3 -r_+ ^2 r_q
\right) \,, \quad &\hbox{for} \quad \omega_f = -2/3 \\
-\frac{\alpha^2}{8 r_+ ^3} \left(2Q^2 - 2r_+ ^2 + 5
\frac{r_+^{7/2}}{r_q ^{3/2}}\right) \,, \quad &\hbox{for} \quad \omega_f = -5/6 
\end{array} \right.
\end{eqnarray}
Replacing these values and the corresponding expressions for the surface
gravity (\ref{kappa}) in Eq.(\ref{Se}) we finally obtain in the three cases,
\begin{equation}
S_\epsilon = \frac{r_+ ^2}{90\alpha^2}\,,
\end{equation}
or in terms of the black hole horizon area $A=4\pi r_+ ^2$,
\begin{equation}
S_\epsilon = \frac{A}{360 \pi \alpha^2} \,.
\end{equation}
This expression is the same correction found by 't Hooft and other
authors for 4-dimensional black holes, a fact that shows its universality. 

\subsection{Thermodynamical stability}

In order to see the influence of the fluid on the stability from a
thermodynamical point of view the first step is to analyze the
specific heat, 
\be
C = T\left(\frac{\partial S}{\partial T}\right)\,,
\ee
which in our case becomes
\be
C = \frac{2\pi \left(3c\,\omega_f r_+ ^2 -  Q^2 r_+ ^{3\omega_f +1} +
  r_+ ^{3\omega_f +3}\right)}{3Q^2 r_+ ^{3\omega_f -1} - r_+
  ^{3\omega_f +1} - 9 c\, \omega_f ^2 - 6 c \, \omega_f} \,.
\ee

\begin{figure}[htp!]
\begin{center}
\includegraphics[height=6.5cm]{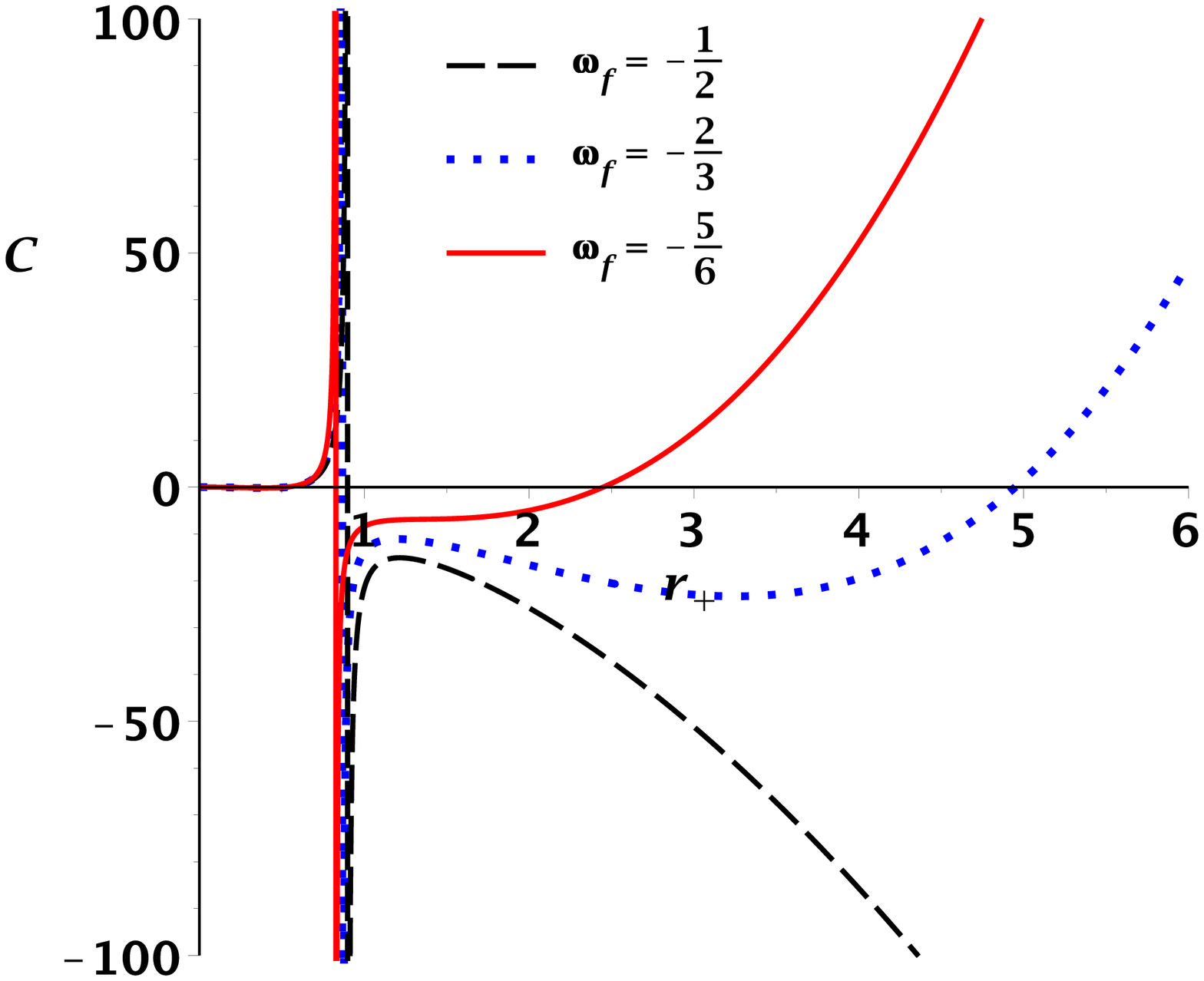}\;
\includegraphics[height=6.5cm]{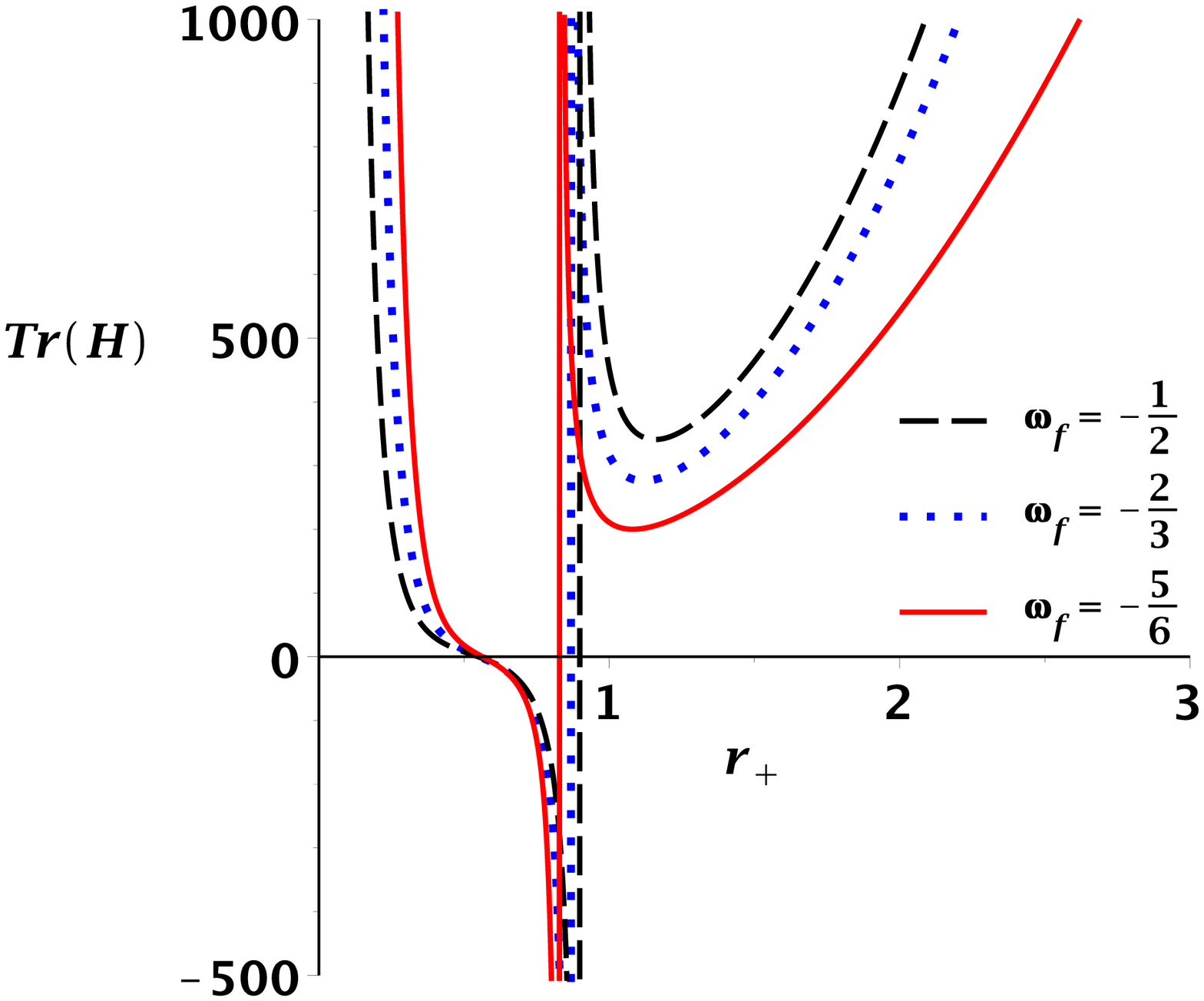}
\caption{Specific heat (left) and trace of Hessian matrix (right) in
  terms of the event horizon for the different values of $\omega_f$
  used in this paper. We set black hole parameters $Q=1/2$ and $c=0.1$.} 
\label{figx}
\end{center}
\end{figure}

\begin{figure}[htp!]
\begin{center}
\includegraphics[height=6.5cm]{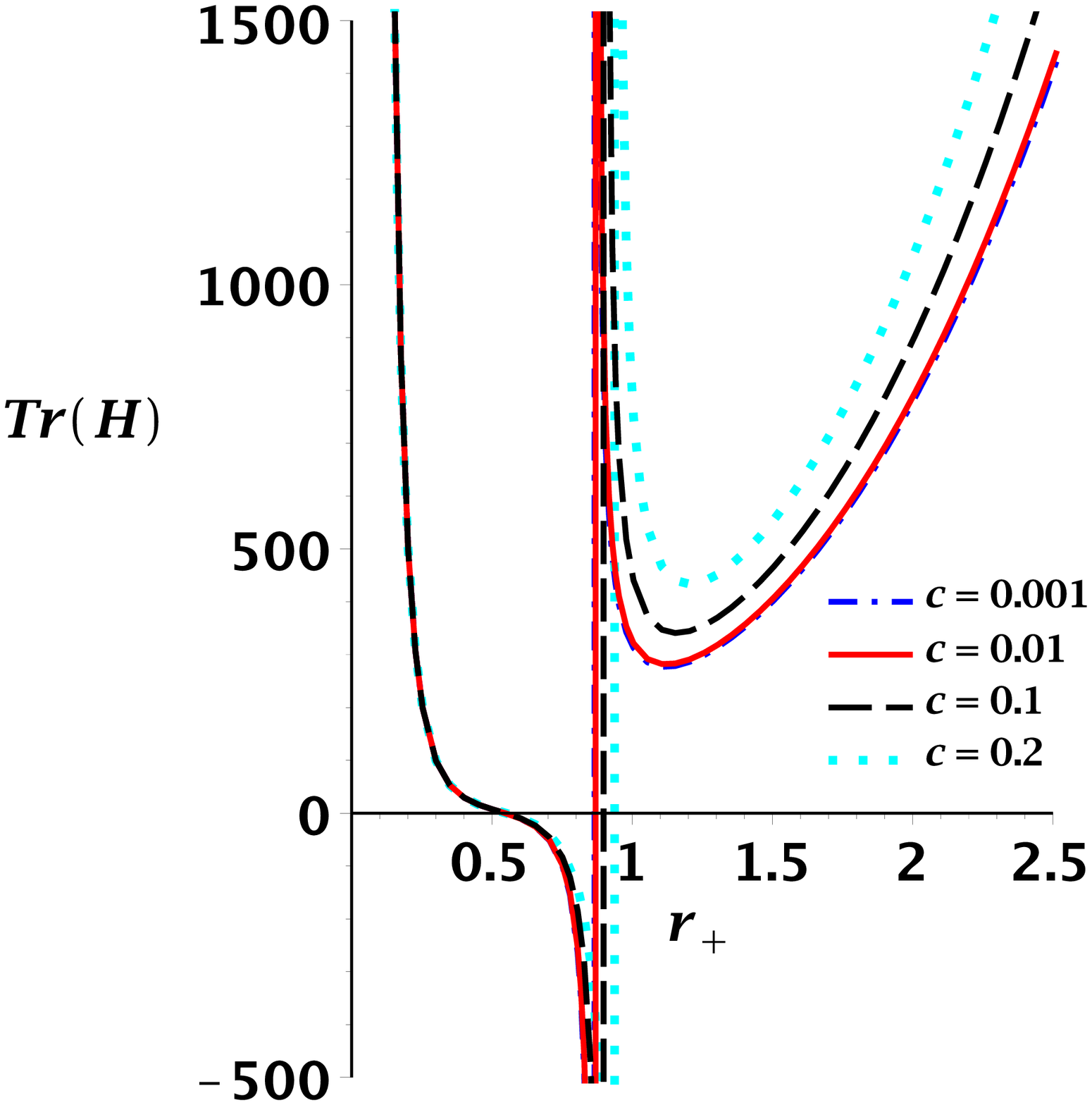}\;
\includegraphics[height=6.5cm]{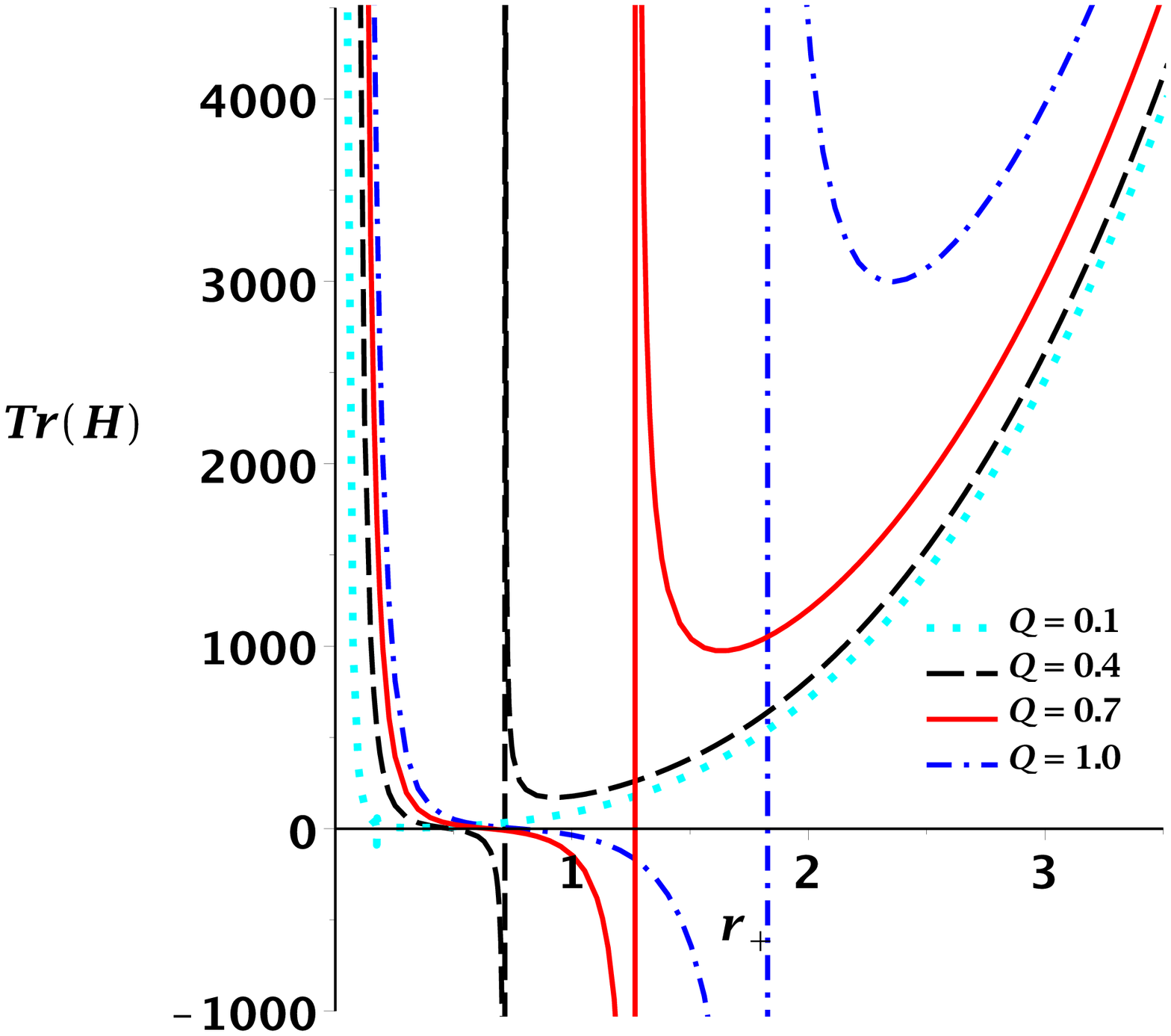}
\caption{Trace of Hessian matrix in terms of the event horizon with
  $Q=1/2$ for different values of $c$ (left) and with $c=0.1$ for
  different values of $Q$ (right).} 
\label{figz}
\end{center}
\end{figure}
The plot of the specific heat for different values of $\omega_f$,
displayed in the left panel of Fig.\ref{figx}, shows the rich
structure of the geometry already noticed in the
literature~\cite{thomas,ghaderi,rodrigue,toledo,saheb}. There are
positive (stable) and negative (unstable) regions alternating with
each other. These regions are separated by several points that signal
first order phase transitions where $C=0$ and also second order
transitions whenever $C$ becomes infinite. However, the sign of the
specific heat is not enough to ensure stability. One additional
criterion to verify the existence of critical points comes from the
Hessian matrix of the Helmholtz free energy ${\cal F}$ related to the the black
hole~\cite{upad}   
\begin{equation}\label{hm}
H = 
\begin{pmatrix}
\frac{\partial ^2 {\cal F}}{\partial T^2} & \frac{\partial ^2 {\cal
    F}}{\partial T \partial {\cal C}} \\
\frac{\partial ^2 {\cal F}}{\partial {\cal C} \partial T} &
\frac{\partial ^2 {\cal F}}{\partial {\cal C}^2}
\end{pmatrix}\,,
\end{equation}
where ${\cal C}$ is the conjugate quantity to the ``charge'' $c$ related
to the presence of the anisotropic fluid given by
\be
{\cal C} = \frac{\partial M}{\partial c} = -\frac{r_+ ^{-3\omega_f}}{2}\,.
\ee
Using the entropy $S=\pi r_+ ^2$ and the temperature of the black hole
$T=\kappa /2\pi$ with $\kappa$ given in Eq.(\ref{kappa}) we find that
\be 
{\cal F} = -\int S \,dT = \frac{r_+}{4} + \frac{3Q^2}{4r_+} -
\left(\frac{1}{2} + \frac{3}{4}\omega_f \right)\,c\,r_+ ^{-3\omega_f}\,.
\ee
With all this information we can calculate the determinant of the
Hessian matrix. However, this determinant vanishes, what means that
one of the eigenvalues of the matrix is zero. The other eigenvalue
corresponds to the trace $\hbox{Tr}(H)$ of the Hessian matrix
(\ref{hm}). Then, a necessary criterion for the model to be stable is
the positivity of this quantity, {\it i.e.}, $\hbox{Tr} (H)\geq 0$. We
plotted this trace in the right panel of Fig.\ref{figx}. We observe
that, in fact, there are regions where $\hbox{Tr} (H) \geq 0$ for the
values of $\omega_f$ considered along this work. Moreover, in the left
panel of Fig.\ref{figz} we see that small black holes fulfill the
stability criterion independent of the value of $c$, whose influence
is only visible for bigger $r_+$. A curious fact is that for $\omega_f
=-2/3$ the trace of the Hessian matrix does not depend on $c$. In
addition, the
effect of the charge on the stability criterion can be seen in the
right panel of Fig.\ref{figz}, small charge black holes have shorter
regions of instability. Therefore, with this analysis we see that it
is possible to have phase transitions for different values of the
black hole and anisotropic fluid parameters.

\section{Final Remarks}\label{sec5}

In this paper we investigate charged black hole spacetimes surrounded
by anisotropic fluids. We firstly obtained the conformal structure of
the entire manifold showing that its Penrose-Carter diagram is similar
to Reissner-Nordstr\"om-dS spacetime, {\it i.e.}, there is a
cosmological-like horizon, an event horizon, and inner Cauchy
horizon. In addition, there is a time-like singularity at the origin
that could be avoided by an observer crossing the inner horizon. The
novelty in the spacetimes considered in the present work is the
light-like structure beyond the cosmological-like horizon
differently from the RN-dS black hole where this region presents
a space-like structure.   

Having established the causal structure of the black hole spacetime we
evolve the scalar field between the event and cosmological-like
horizons obtaining two interesting features. The first one is that the
late-time behavior is dominated firstly by a power-law term for small
state parameter of the fluid $|\omega_f|$ and, afterwards, by an
exponential decay (purely imaginary quasi normal mode) for higher
$|\omega_f|$. For these geometries the presence of a power-law term in
the final stage comes as an interesting new result never reported
before even in de Sitter black hole spacetimes where this phenomenon
is also present. The second one concerns the quasinormal modes
obtained in Section \ref{sec3}. They provide the spectrum of
oscillation of the black hole when perturbed by a scalar field. We
show that they are very similar for different state parameters when
the fluid density is small being hugely influenced when it becomes
large. When varying the state parameter, the oscillations have both
imaginary and real part diminished as we increase $|\omega_f|$. 

Regarding the thermodynamical calculations, we considered an arbitrary
particle of proper energy $E$ in equatorial motion and captured by these
black holes surrounded by anisotropic fluids. Our result shows that
these geometries yield the universal bound for the entropy of the
falling system originally found by
Bekenstein~\cite{PhysRevD.23.287}. In addition, we also considered a
thermal bath of scalar fields propagating outside the event horizon of
these black holes in order to find the semiclassical corrections to
their entropy. Following 't Hooft's brickwall method we found the same
kind of correction corresponding to 4-dimensional black holes showing
the universality of this result~\cite{tHooft:1984kcu}. Finally, we
also analyzed the thermodynamical stability looking at the specific
heat of the black hole. As an additional criterion to ensure the
presence of critical points, we also calculated the trace of the
Hessian matrix of the Helmholtz free energy. In this way we showed that
phase transitions of first and second order are possible for different
values of the black hole and anisotropic fluid parameters.

\begin{acknowledgments}
The authors would like to thank Matt Visser and Christos Charmousis
for useful discussions. This work was partially supported by UFMT ({\it {Universidade Federal de Mato Grosso}}) under Grant PROPG $005/2019$.
\end{acknowledgments}

\appendix

\section{Maximal extension for the black hole solution with $w_f=-2/3$}\label{apendiceB}

The case with $w_f=-2/3$ has the following line-element
\begin{equation}\label{wf_2_3}
ds^{2}=-f(r)dt^2 +f(r)^{-1}dr^2 + r^2 d\Omega^{2}_2,
\end{equation}
with
\begin{equation}\label{f}
f(r)= 1-\frac{2M}{r}+\frac{Q^2}{r^2}-cr .
\end{equation}

In the cases where $M>Q$ it is possible to express the metric components $g_{tt}$ and $g_{rr}$ in terms of three distinct positive real roots $(r_c, r_{+}, r_{-})$ which, as in the case $w_f=-1/2$, represents the cosmological-like horizon, event horizon, and Cauchy horizon, respectively. So,
\begin{equation}\label{f_roots}
f(r)=-\frac{c}{r^2}(r-r_c)(r-r_{+})(r-r_{-}),
\end{equation}
and the tortoise coordinate $r_{*}$ can be written as
\begin{equation}
r_{*}=-\frac{1}{\kappa_c}\log{|r-r_c|}+\frac{1}{2\kappa_+}\log{|r-r_+|}-\frac{1}{\kappa_-}\log{|r-r_-|},
\end{equation}
with $(\kappa_c,\kappa_+,\kappa_-)$ referring to the surface gravity in each horizon. Following the same steps as in the case $w_f=-1/2$, we obtain the maximal extension in each horizon. For the cosmological-like horizon $r=r_c$ we have
\begin{equation}\label{cosmo_2_3}
U_cV_c = \pm |r-r_c|\frac{1}{|r-r_{+}|^{\kappa_c/\kappa_+}}|r-r_-|^{\kappa_c/\kappa_-},
\end{equation}
where the upper sign refers to $r>r_c$ and the lower sign corresponds to $r<r_c$. In the cases of event horizon $r_+$ and Cauchy horizon $r_-$, we have found similar expressions,
\begin{equation}\label{event_2_3}
U_+V_+ = \mp |r-r_+|\frac{1}{|r-r_c|^{\kappa_+/\kappa_c}}\frac{1}{|r-r_-|^{\kappa_+/\kappa_-}},
\end{equation}
and
\begin{equation}\label{cauchy_2_3}
U_-V_-=\pm|r-r_-||r-r_c|^{\kappa_-/\kappa_c}\frac{1}{|r-r_+|^{\kappa_-/\kappa_+}}.
\end{equation}
Thus, introducing the Penrose coordinates $T=\frac{1}{2}(\tilde{V}+\tilde{U})$ and $R=\frac{1}{2}(\tilde{V}-\tilde{U})$ in each horizon we obtain the Penrose-Carter diagram as shown in Fig.\ref{penrose_1_2}.



\end{document}